\documentclass[apj]{emulateapj} 

\usepackage{epsfig}
\usepackage{amsmath}
\usepackage{amssymb}
\usepackage{natbib}
\usepackage{threeparttable} 
\usepackage{booktabs}

\usepackage{epstopdf}

\newcommand{\chan}{\textit{Chandra}}
\newcommand{\swift}{\textit{Swift}}

\newcommand{\xmm}{\textit{XMM-Newton}}

\newcommand{\maxi}{\textit{MAXI}}

\newcommand{\smarts}{\textit{SMARTS}}

\newcommand{\Msun}{\mathrm{M}_{\odot}}

\newcommand{\flux}{\mathrm{erg~cm}^{-2}~\mathrm{s}^{-1}}

\newcommand{\cnts}{\mathrm{counts~s}^{-1}}

\newcommand{\nh}{\mathrm{cm}^{-2}}

\newcommand{\source}{Swift J1910.2--0546}
\newcommand{\maxiname}{MAXI J1910--057}
\newcommand{\xte}{XTE J1817--330}

\newcommand{\bhmontse}{Swift J1357.2--0933}
\newcommand{\gx}{GX 339--4}

\def \atel {ATel}
\def \na {New Astronomy}
\slugcomment{Received 2013 November 28; accepted 2014 February 21}

\shorttitle{Black Hole Candidate \source}
\shortauthors{Degenaar et al.}

\begin{document}

\title{Multi-Wavelength Coverage of State Transitions in the New Black Hole X-Ray Binary Swift J1910.2--0546}

\author{N. Degenaar$^{1,}$\altaffilmark{8}, D. Maitra$^{1,2}$, E.M. Cackett$^{3}$, M.T. Reynolds$^{1}$, J.M. Miller$^{1}$, R.C. Reis$^{1,9}$, A.L. King$^{1}$, K.G\"ultekin$^{1}$, C.D.~Bailyn$^{4}$, M.M. Buxton$^{4}$, R.K.D. MacDonald$^{4}$, A.C. Fabian$^{5}$, D.B. Fox$^{6}$, E.S. Rykoff$^{7}$
}
\affil{$^1$Department of Astronomy, University of Michigan, 500 Church Street, Ann Arbor, MI 48109, USA; degenaar@umich.edu\\
$^2$Department of Physics and Astronomy, Wheaton College, 26 East Main Street, Norton, MA 02766, USA\\
$^3$Department of Physics and Astronomy, Wayne State University, 666 West Hancock Street, Detroit, MI 48201, USA\\
$^4$Astronomy Department, Yale University, P.O. Box 208101, New Haven, CT 06520-8101, USA\\
$^5$Institute of Astronomy, University of Cambridge, Madingley Road, Cambridge CB3 OHA, UK\\
$^6$Department of Astronomy and Astrophysics, Pennsylvania State University, 525 Davey Laboratory, University Park, PA 16802, USA\\
$^7$SLAC National Accelerator Laboratory, Menlo Park, CA 94025, USA
}
\altaffiltext{8}{Hubble Fellow}
\altaffiltext{9}{Einstein Fellow}

\begin{abstract}
Understanding how black holes accrete and supply feedback to their environment is one of the outstanding challenges of modern astrophysics. \source\ is a candidate black hole low-mass X-ray binary that was discovered in 2012 when it entered an accretion outburst. To investigate the binary configuration and the accretion morphology we monitored the evolution of the outburst for $\simeq$3 months at X-ray, UV, optical ($B,V,R,I$), and near-infrared ($J,H,K$) wavelengths using \swift\ and \smarts. The source evolved from a hard to a soft X-ray spectral state with a relatively cold accretion disk that peaked at $\simeq$0.5~keV. A \chan/HETG spectrum obtained during this soft state did not reveal signatures of an ionized disk wind. Both the low disk temperature and the absence of a detectable wind could indicate that the system is viewed at relatively low inclination. The multi-wavelength light curves revealed two notable features that appear to be related to X-ray state changes. Firstly, a prominent flux decrease was observed in all wavebands $\simeq1-2$ weeks before the source entered the soft state. This dip occurred in (0.6--10 keV) X-rays $\simeq6$~days later than at longer wavelengths, which could possibly reflect the viscous time scale of the disk. Secondly, about two weeks after the source transitioned back into the hard state, the UV emission significantly increased while the X-rays steadily decayed. We discuss how these observations may reflect changes in the accretion morphology, perhaps related to the quenching/launch of a jet or the collapse/recovery of a hot flow. 
\end{abstract}


\section{Introduction}
Black holes in low-mass X-ray binaries (LMXBs) accrete matter from a low-mass ($\lesssim$1$~\Msun$) companion star that overflows its Roche lobe. During accretion outbursts, matter is supplied to the black hole through an accretion disk and outflows in the form of disk winds and jets are generated. X-ray monitoring has revealed a rich timing and spectral behavior leading to the formulation of distinct X-ray spectral states and transitions between them \citep[e.g.,][]{homan2005_specstates,remillard2006}.\footnote[10]{In this work we discuss X-ray spectral data in the 0.6--10 keV energy range and therefore we adopt a simplified soft/hard state nomenclature that neglects intermediate states.}

During {\it soft} states, the X-ray spectrum is dominated by thermal emission from the hot inner part of the accretion disk with a typical temperature of $kT_{\mathrm{in}}\simeq$1~keV \citep[e.g.,][]{dunn2011,reynolds2013}. Strong ionized X-ray disk winds are commonly detected during this state \citep[e.g.,][]{miller2006_winds,ponti2012_winds}. In {\it hard} states, the accretion disk is colder and the X-ray spectrum is dominated by a power law with a photon index of $\Gamma \simeq 1.5-2$. Disk winds seem to be suppressed \citep[e.g.,][]{miller2008,miller2012_winds,neilsen2009}, and instead compact radio jets are observed \citep[e.g.,][]{fender2005}. The hard X-rays are ascribed to a hot disk corona, a radiatively inefficient accretion flow, or a jet \citep[e.g.,][]{esin1997,markoff2001,brocksopp2004}.  

There are several mechanisms that are thought to contribute to the optical, ultraviolet (UV) and near-infrared (nIR) emission of black hole LMXBs. This radiation may originate in the cool outer parts of the accretion disk as the result of viscous dissipation or reprocessing of X-rays, but could also be produced in the jet, corona or inner hot flow \citep[e.g.,][]{paradijs95,esin1997,markoff2001,russell2006,rykoff2007,veledina2011}. The late-type companion star is typically dim and not contributing significantly to the outburst flux.

Multi-wavelength observations of the outburst evolution can shed light on the physical mechanisms producing the different emission components, the accretion inflow-outflow coupling, and the morphological changes occurring during X-ray state transitions. In this work we report on such a study of the newly discovered candidate black hole LMXB \source.

\section{Swift J1910.2--0546}\label{subsec:source}
\source\ (\maxiname) is a transient X-ray source that was discovered on 2012 May 30--31 through the monitoring surveys of \swift/BAT and \maxi\ \citep[][]{krimm2012,usui2012}. An optical/nIR counterpart was readily identified and indicated an LMXB nature \citep[][]{rau2012_1910,kennea2012_1910}. Initially, the optical emission was found to be highly variable, suggestive of a relatively short orbital period of $\simeq$2--4~hr \citep{lloyd2012}, but later spectroscopic studies favored a wider orbit of $\gtrsim6.2$~hr \citep[][]{casares2012}. Apart from these estimates, no orbital ephemeris has been reported at present.

X-ray spectral analysis showed that the source traced out the canonical hard and soft states \citep[][]{kennea2012_1910,kimura2012,nakahira2012,bodaghee2012,reis2013}, and a steady radio jet was detected \citep[][]{king2012_1910}. \source\ is therefore considered a new candidate black hole LMXB. X-ray reflection features detected with \xmm\ suggested that the inner accretion disk may be truncated at a higher luminosity (i.e., a higher mass-accretion rate) than is typically seen in other black hole LMXBs, or alternatively could point to a retrograde spinning black hole \citep[][]{reis2013}.

\section{Observations and Data Analysis}
\source\ is detected in the \swift/BAT and \maxi\ monitoring surveys for $\simeq$35 weeks starting around MJD 56077 (2012 May 30).\footnote[11]{The \swift/BAT and \maxi\ light curves are consistent with the background level from 2013 January/February onward, although pointed \swift/XRT and ATCA radio observations indicated low-level accretion activity in 2013 May \citep[][]{tomsick2013}.} We obtained contemporaneous X-ray, UV, optical and nIR monitoring observations between MJD 56140 and 56254 (2012 August 1 till November 23), using the \swift\ and \smarts\ observatories. These observations sample $\simeq$3~months of the full $\simeq$9-month outburst. 
We also obtained a \chan\ observation to search for signatures of an ionized disk wind.

\subsection{\chan\ HETG Observation}\label{subsec:chan}
\source\ was observed with \chan\ for $\simeq$30~ks starting on 2012 September 22 at UT 23:22 (MJD 56192, ObsID 14634). The incident flux was dispersed onto the ACIS-S CCDs using the High Energy Transmission Grating (HETG) with the ACIS-S array operated in the ``CC33$\_$GRADED" mode. The data were processed using the \textsc{ciao} software (ver. 4.4). The positive and negative diffraction orders of the High-Energy Grating (HEG) and Medium-Energy Grating (MEG) spectra were combined using \textsc{add\_grating\_orders}. The spectral data were fitted in \textsc{XSpec} \citep[ver. 12.7;][]{xspec}.

\begin{figure}
 \begin{center}
\includegraphics[width=8.5cm]{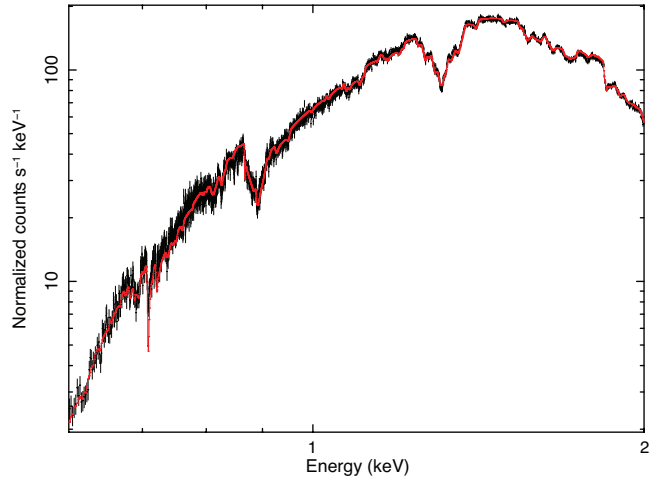}
    \end{center}
\caption[]{{\chan/MEG spectrum obtained during the soft state on MJD 56192 (2012 September 22). The red curve shows a fit using an absorbed disk black body. There are no X-ray absorption features that would testify to the presence of an ionized disk wind.}}
 \label{fig:meg}
\end{figure}

\subsection{\swift\ XRT and UVOT Observations}\label{subsec:swift}
\source\ was monitored during 47 \swift\ observations of $\simeq$1~ks that were performed between 2012 August 7 and November 23 (target ID 32521). These provide simultaneous coverage in X-rays and UV via the X-Ray Telescope (XRT) and the Ultra-Violet/Optical Telescope (UVOT), respectively. The \swift\ data were reduced using \textsc{heasoft} (ver. 6.13).

All XRT observations were carried out in Windowed Timing mode and processed using the task \textsc{xrt$\_$pipeline}. The source was detected with raw XRT intensities varying between $\simeq$6 and $250~\cnts$. After examining the data for possible pile-up, we chose to conservatively exclude the central 3 pixels for count rates of $\simeq$150--200~$\cnts$ and 5 pixels for $\simeq$200--250$~\cnts$  \citep[see also][]{reynolds2013}. 

Using \textsc{XSelect}, we extracted XRT spectra from a box of $60$ pixels long and $20$ pixels wide centered on the source. Background events were obtained from neighboring regions using a box of the same dimensions. The corresponding arfs were created using the tool \textsc{xrtmkarf}, and the rmfs (ver. 14) were sourced from the calibration data base. The spectral data were grouped to a minimum of 20 photons per bin. We also investigated the spectral evolution by determining the X-ray hardness ratio for each observation. For the purpose of the present work we defined this as the ratio of counts in the 1.5--10 keV and 0.6--1.5 keV energy bands.

All UVOT images were obtained using the $uvm2$ filter ($\lambda_c$$\simeq$$2246$~\AA). Magnitudes and flux densities were extracted with the tool \textsc{uvotsource} using a standard aperture of $5''$. A nearby, source-free aperture of $15''$ served as a background reference. 

\subsection{\smarts\ Optical and nIR Observations}\label{subsec:smarts}
We monitored \source\ at optical and nIR wavelengths using the ANDICAM detector mounted on the \smarts\ 1.3-m telescope at the Cerro Tololo Inter-American Observatory in Chile. Between 2012 August 1 and October 17 the source was observed 42 times using a selection of $B, V, R, I, J, H,$ and $K$ filters. 

Images of 250 s were obtained for the $B$ and $V$ filters, while the $R$ and $I$ band exposures were 300~s. The nIR data consisted of seven dithered images of 40~s each in $J$, seven dithered images of 40~s each in $H$, and thirteen dithered images of 25~s each in $K$. These dithered frames were flat-fielded, sky subtracted, aligned, and average-combined using an in-house \textsc{iraf} script. 

Two nearby 2MASS stars were taken as references to perform optical and nIR differential photometry with respect to \source. Their optical magnitudes were taken from the NOMAD and USNO-B1 catalogs. The night-to-night variability in the reference stars is accounted for in the magnitude errors.

\section{Results}\label{sec:results}
\subsection{No Sign of an Ionized Disk Wind}\label{subsec:nh}
The \chan/HETG observation was obtained during a soft state (see Section~\ref{subsec:obevo}). However, the HEG and MEG spectra do not show X-ray absorption lines that would evidence the presence of an ionized disk wind (see Figure~\ref{fig:meg}). We fitted the MEG data between 0.5 and 2.0 keV using an absorbed disk black body \citep[\textsc{diskbb};][]{mitsuda1984} to constrain the hydrogen column density. For this purpose we adopted the \textsc{tbnew} prescription with the \textsc{vern} cross-sections and \textsc{wilm} abundances \citep[][]{verner1996,wilms2000}.

The spectral fit shown in Figure~\ref{fig:meg} yielded a hydrogen column density of $N_{\mathrm{H}} = (3.5 \pm 0.1)\times10^{21}~\nh$, and a disk temperature of $kT_{\mathrm{in}} = 0.42\pm0.01$~keV. Allowing the abundances of O, Ne, Mg and Fe to vary did not indicate any prominent deviations from Solar composition.

\begin{figure}
 \begin{center}
\includegraphics[width=8.7cm]{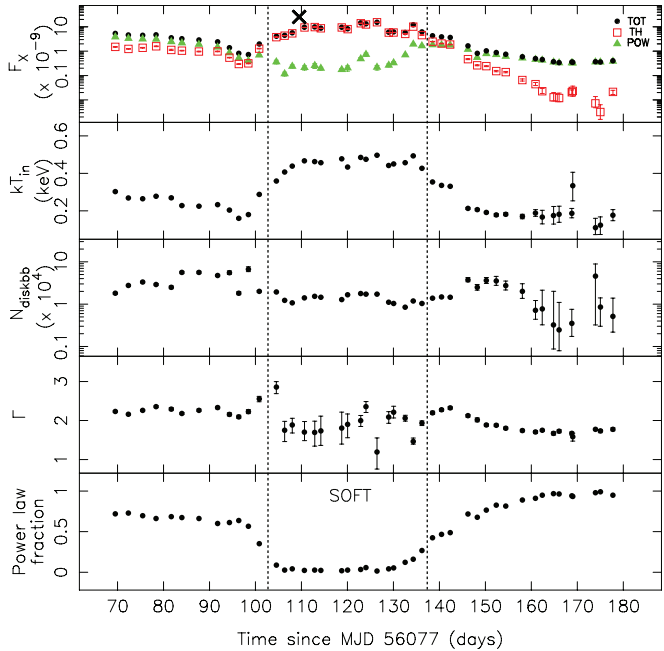}
    \end{center}
\caption[]{{X-ray spectral evolution. From top to bottom: The 0.6--10 keV unabsorbed flux in units of $10^{-9}~\flux$ (red squares for the disk, green triangles for the power law and black filled circles for the total flux), the inner disk temperature, the disk normalization, the power-law index, and the fractional contribution of the power law to the total flux. The ``X'' in the top panel marks the time of our \chan/HETG observation. The vertical dotted lines illustrate when the source was in a soft state. Error bars represent $1\sigma$ confidence levels.}}
 \label{fig:swift}
\end{figure}

\begin{figure}
 \begin{center}
\includegraphics[width=8.7cm]{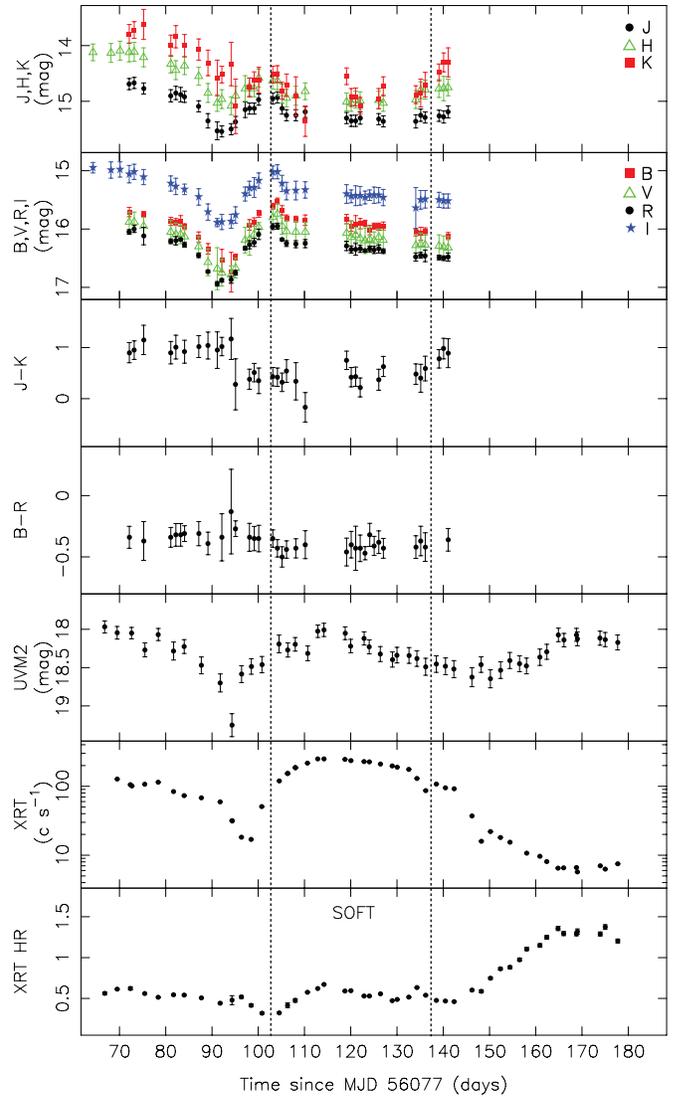}
    \end{center}
\caption[]{{Multi-wavelength light curves and color evolution. From top to bottom: nIR, optical, nIR color, optical color, UV, 0.6--10 keV X-rays and X-ray hardness ratio (1.5--10 keV / 0.6--1.5 keV). The vertical dotted lines mark the time at which the source was in a soft state.}}
 \label{fig:lc}
\end{figure}

\subsection{State Transitions and a Cool Disk}\label{subsec:obevo}
To obtain X-ray fluxes and to characterize the outburst evolution, all XRT spectra were fitted to a combined disk black body (\textsc{diskbb}) and a power law (\textsc{powerlaw}). Interstellar absorption was taken into account by using the \textsc{tbabs} model with the hydrogen column density fixed to $N_{\mathrm{H}} = 3.5\times10^{21}~\nh$ \citep[Section~\ref{subsec:nh}; for a justification, see][]{miller2009}. X-ray fluxes were calculated in the 0.6--10 keV range. The results of our spectral analysis are illustrated in Figure~\ref{fig:swift}.

Our \swift\ observations started on 2012 August 7 (MJD 56146), shortly after \source\ had entered a hard state \citep[][]{nakahira2012,bodaghee2012,king2012_1910}. The XRT spectra were initially dominated by a hard power law with an index of $\Gamma \simeq 2$, but thermal emission from a relatively cool accretion disk with a temperature of $kT_{\mathrm{in}}\simeq0.25$~keV contributed $\simeq$40\% to the total 0.6--10 keV flux (Figure~\ref{fig:swift}).\footnote[12]{The relatively soft photon index and low hardness ratio suggests that the source may have been in an intermediate hard state at this time (see Figures~\ref{fig:swift} and~\ref{fig:lc}).}

Around September 5 (MJD 56175), the X-ray spectrum softened considerably. The disk temperature increased and the thermal emission dominated the 0.6--10 keV flux from September 10 onward (MJD 56180), indicating that \source\ had entered a soft state (Figure~\ref{fig:swift}).\footnote[13]{The power-law index was not well constrained during the soft state because the hard flux contributed only $\simeq$1\% to the total 0.6--10 keV flux.} Notably, even at the peak of the soft state when the thermal emission accounted for $\simeq$99\% of the total flux, the disk remained relatively cool with $kT_{\mathrm{in}}\simeq0.5$~keV.
 
After October 2 (MJD 56202), the thermal flux started to decrease while the non-thermal flux increased. The disk cooled to a temperature of $kT_{\mathrm{in}}\simeq0.2$~keV and the power law hardened to $\Gamma \simeq 1.8$.\footnote[14]{Even in the hardest spectra the addition of a soft component improves the fit over a single power law. For example, for the last observation (ObsID 32521049) adding a disk black body improves the fit from $\chi^2=274.93$/236 dof to $\chi^2=256.92$/234 dof with an $f$-test probability of $\simeq3\times10^{-4}$.} During the last phase of our monitoring campaign (October 25 till November 23, MJD 56225--56254), the X-ray spectrum is completely dominated by the hard power law, demonstrating that the source was back in a hard state (Figure~\ref{fig:swift}). \source\ thus traced out the typical black hole X-ray spectral states \citep[see also][]{reis2013}, and made two state transitions during our monitoring campaign.

\subsection{Features in the Multi-wavelength Light Curves}\label{subsec:dip}
Figure~\ref{fig:lc} displays the nIR, optical, UV and 0.6--10 keV X-ray light curves of \source, as well as the evolution of the nIR ($J$--$K$) and optical ($B$--$R$) colors, and the X-ray hardness ratio. A striking feature is a dip in intensity that occurs $\simeq$95 days into the outburst. At all wavelengths the flux steadily decreases toward a minimum on a time scale of $\simeq$10 days, after which it recovers on a similar time scale. The optical and nIR magnitudes faded by $\simeq0.9$~mag, which corresponds to a $\simeq$55\% drop in flux. The UV magnitude faded by $\simeq$1.3~mag, or $\simeq$70\%  in flux. The soft (disk) and hard (power law) X-ray flux decreased by $\simeq$80\% and 90\%, respectively. 

\begin{figure}
 \begin{center}
\includegraphics[width=8.7cm]{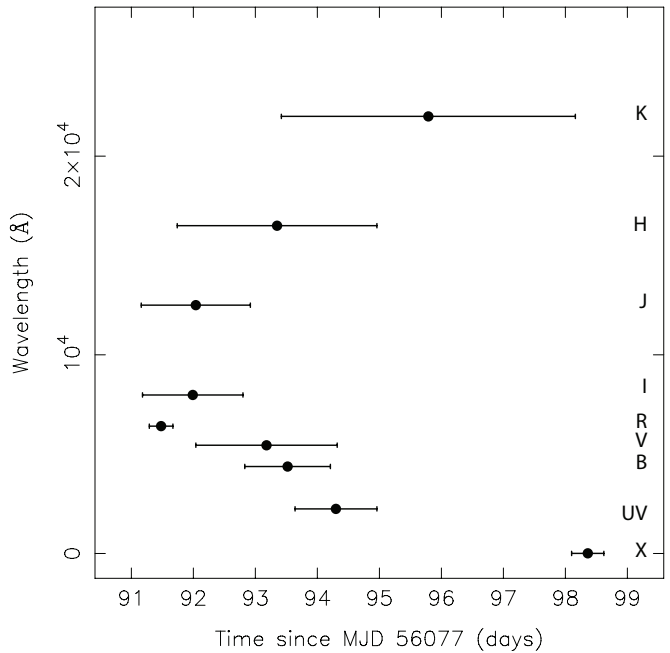}
    \end{center}
\caption[]{{Time of minimum flux during the dip between MJD 56162--56182 (days 85--105) as measured in the different wavebands (indicated on the right).
}}
 \label{fig:dip}
\end{figure}

The measured X-ray flux reached a minimum on 2012 September 5 (MJD 56175; ObsID 32521012), whereas the point of lowest UV flux was recorded two observations earlier on September 1 (MJD 56171; ObsID 32521010). This is suggestive of a time delay between the different wavelengths \citep[see also][]{krimm2013}. In an attempt to quantify this delay, we fitted the light curves to a simple broken exponential to find the time of minimum flux in each band between August 23 and September 12 (days 85--105 in Figure~\ref{fig:lc}). The results are listed in Table~\ref{tab:lags} and illustrated in Figure~\ref{fig:dip}. Although the error margins are large, it appears that the dip first occurred in $R$, $I$, and $J$ around day 92 (MJD 56169), whereas both longer ($H$ and $K$) and shorter ($B$, $V$, and $UV$) wavelengths are delayed by $\simeq1-4$~days, followed by X-rays 6 days later (around day 98).

We also searched for and quantified any time lags using the standard cross-correlation analysis techniques widely used for optical reverberation in active galactic nuclei.  We used a linear-interpolation cross correlation \citep[][]{white1994} and determined the uncertainties in the lags using a flux randomization and random subset sampling Monte Carlo method \citep{peterson1998,peterson2004}. There were not enough points to apply this technique to a subset of the data covering the time of the flux dip only. Assuming that the variability in each waveband and each spectral state are caused by the same (or otherwise different but causally related) physical processes, we applied the method to the full data set. This analysis suggests that in the overall light curves there is a negligible time difference between the optical and nIR bands, but that the UV and X-rays lag the longer wavelengths by $\simeq$3 and $8$~days, respectively. These lags are not exactly the same as those estimated from the flux dip only (see Table~\ref{tab:lags}), which may be the result of different emission mechanisms operating in the hard and soft spectral states. 

Notably, \source\ entered a soft state $\simeq$1 week after the intensity dip was seen in the X-rays (Section~\ref{subsec:obevo}). This suggests a possible causal connection between the flux dip and the state transition (see Section~\ref{sec:discussion}). This is corroborated by the fact that around the time of the flux dip the nIR color suddenly became much bluer, e.g., the $J-K$ color abruptly dropped from $\simeq$1 to 0.5 around day 95 in Figure~\ref{fig:lc}. In contrast, the optical bands continued to vary together, with little change in color, as illustrated by the evolution of $B-R$. The nIR colors remained blue during the soft state, only to become redder again when the $H$ and $K$ bands brightened  and the source transitioned back into a hard state (around day 138; see also Section~\ref{subsec:flow}).

Whereas the \smarts\ optical/nIR observations stopped at the end of the soft state, X-ray/UV monitoring with \swift\ continued. This revealed that the X-rays steadily decayed into the hard state, whereas the UV emission started to rise $\simeq2$~weeks after the soft to hard state transition (around day 158). At the same time the X-ray spectrum considerably hardened, as illustrated by the strong increase in XRT hardness ratio (Figure~\ref{fig:lc}).

\begin{table}
\begin{center}
\caption{Time Lags in \source.\label{tab:lags}}
\begin{tabular*}{0.49\textwidth}{@{\extracolsep{\fill}}ccc}
\hline
\hline
Waveband & \multicolumn{2}{c}{Lag}\\
 & \multicolumn{2}{c}{(days)}\\
\hline
& Flux Dip & Full Light Curve \\
UV$-$X-ray & $4.06\pm0.71$ & $7.72^{+0.80}_{-0.71}$\\
$B -$X-ray & $4.84\pm0.74$ & $8.24\pm0.43$\\
$B -$UV & $0.78\pm0.95$ & $2.72^{+1.00}_{-0.58}$\\
$B - V$ & $-0.34\pm1.33$ & $-0.13\pm0.26$\\
$B - R$ & $-2.04\pm0.72$ & $-0.52\pm0.18$\\
$B - I$ & $-1.53\pm1.06$ & $-0.52\pm0.22$\\
$B - J$ & $-1.48\pm1.12$ & $-0.38^{+0.28}_{-0.24}$ \\
$B - H$ & $-0.17\pm1.75$ & $-0.13^{+0.49}_{-0.44}$\\
$B - K$ & $2.27\pm2.47$ & \nodata \\
\hline
\end{tabular*}
\tablecomments{The waveband listed first is taken as the reference, i.e., the lag given is with respect to that. When considering the full light curve no lag could be determined for the $K$ band, as it was poorly correlated with the other light curves. }
\end{center}
\end{table}

\subsection{Relation between the X-Ray and UV Flux}\label{subsec:uvcorr}
\begin{figure*}
 \begin{center}
\includegraphics[width=8.5cm]{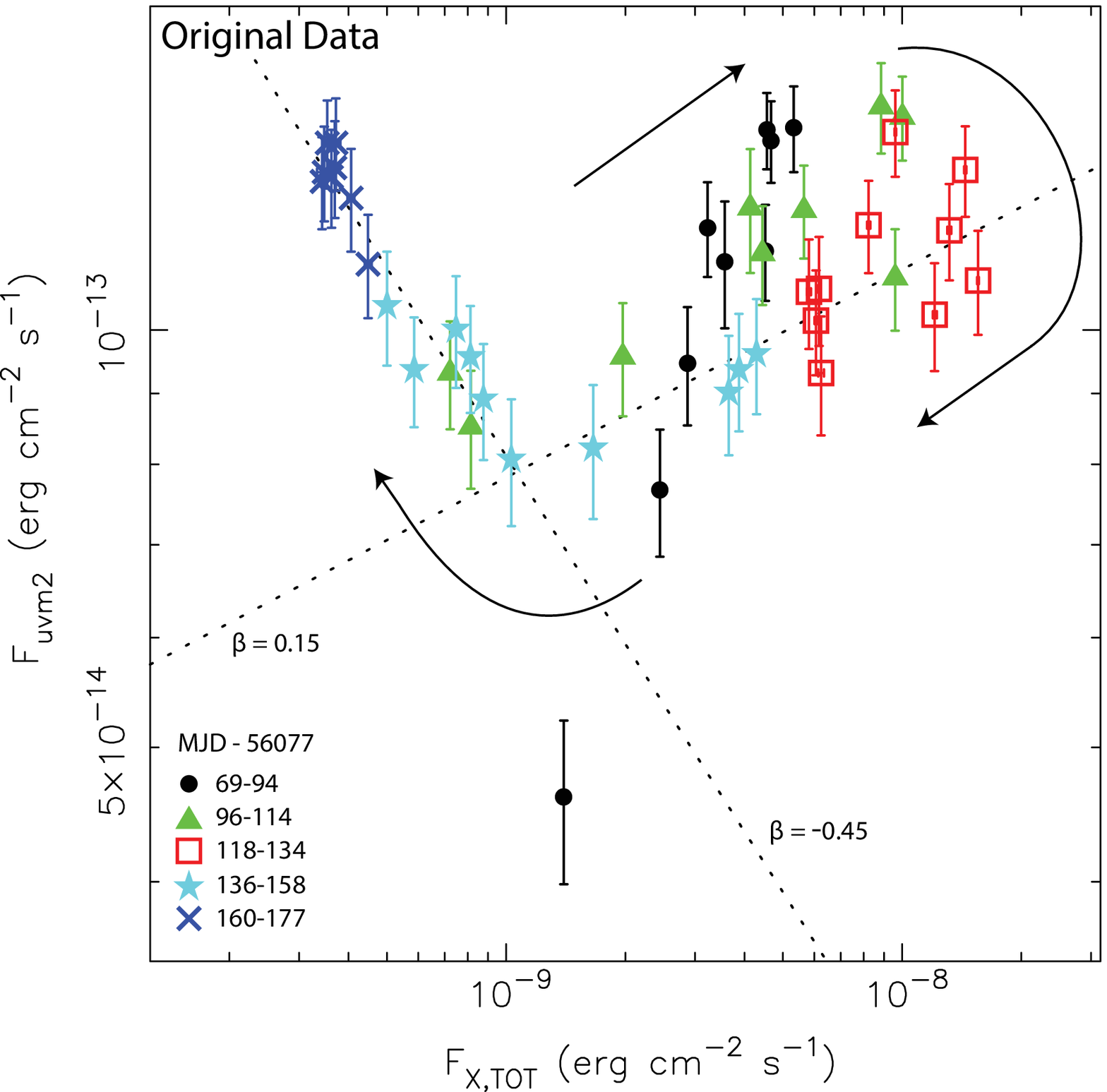}\hspace{+0.5cm}
\includegraphics[width=8.5cm]{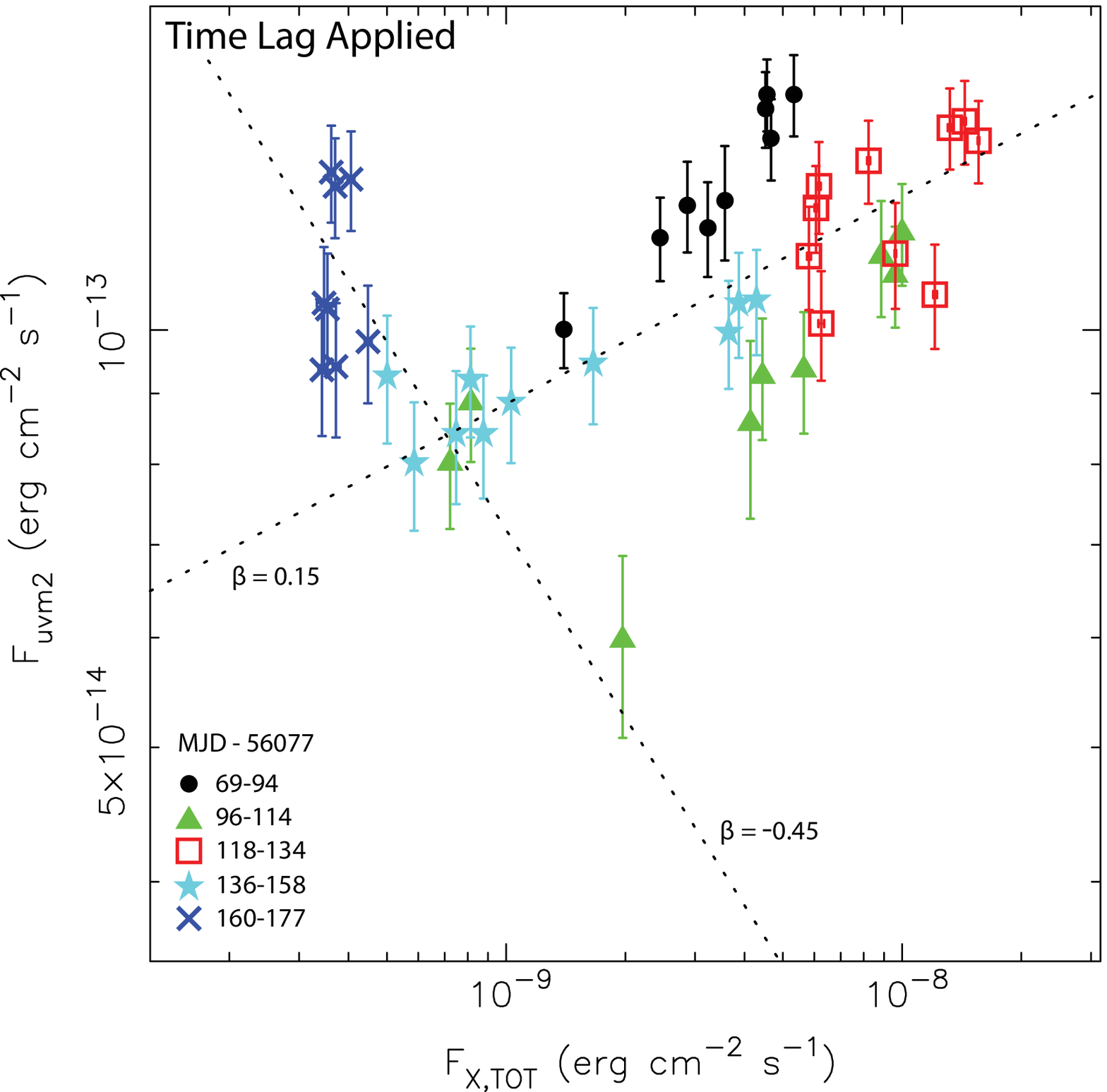}
    \end{center}
\caption[]{{Relation between the \swift\ X-ray and UV fluxes. Dotted lines indicate fits to a power law of the form $F_{\mathrm{uvm2}}\propto F_{X}^{\beta}$. The temporal evolution is encoded in the different symbols, which indicate the time in days since MJD 56077 (arrows are indicated to guide the eye). The right plot illustrates how the relation changes after applying a time delay of 7.72 days for the X-rays with respect to the UV. }}
 \label{fig:uvcorr}
\end{figure*}

To investigate the apparent anti-correlation seen between the UV and X-rays in the hard state (Section~\ref{subsec:dip}), we calculated UV fluxes ($F_{\mathrm{uvm2}}$) by multiplying the measured UVOT flux densities with the FWHM of the $uvm2$ filter \citep[$\simeq$$498$~\AA;][]{poole2008}. In Figure~\ref{fig:uvcorr} we plot the relation between the UV and the 0.6--10 keV X-ray flux. Above $F_X\simeq10^{-9}~\flux$, the two are positively correlated (Spearman rank coefficient $\rho=0.52$, corresponding to a $\simeq$3$\sigma$ significance), whereas at lower fluxes a very significant negative correlation is apparent ($\rho=0.96$; $\simeq$7$\sigma$). 

To quantify these relations, we fitted both sides separately to a power law of the form $F_{\mathrm{uvm2}} \propto F_{X}^{\beta}$. This yielded $\beta = 0.15\pm0.05$ and $-0.45\pm0.03$ for the positive and negative correlation, respectively (dotted lines in Figure~\ref{fig:uvcorr}). A fit with a broken power law yields similar indices ($\beta \simeq 0.22$ and $-0.45$), and a break located at $F_X\simeq1.1\times10^{-9}~\flux$. 

To investigate the time evolution along the X-ray/UV relations, we divided the outburst into five contiguous intervals of 9/10 observations: These are indicated by different symbols in Figure~\ref{fig:uvcorr}. This illustrates that the UV/X-ray relation evolved from a positive into a negative correlation once the source had moved from the soft to the hard state (around day 138). The anti-correlation thus mostly arises during the last part of the light curves when the UV emission rises while the X-rays decrease (Figure~\ref{fig:lc}).

It is plausible that the observed anti-correlation is introduced due to the presence of a time lag between the X-ray and UV emission. To gauge this possibility, we investigated the relation after applying a time lag of 7.72 days (Section~\ref{subsec:dip}). Figure~\ref{fig:uvcorr} (right) shows that a negative correlation arising from the last set of observations persists, although it is less significant ($\rho=0.68$; $\simeq$2$\sigma$). The positive correlation is now stronger ($\rho=0.59$; $\simeq$4$\sigma$). A fit with a broken power law yields the same indices as before applying the time lag ($\beta \simeq 0.15$ and $-0.45$), and a break located at $F_X\simeq7\times10^{-10}~\flux$. 

The observed (anti-) correlation is thus sensitive to the presence of a time lag. We determined a 7.72 days time delay between the X-ray and UV emission using the full light curves. However, since different emission mechanisms could be contributing to the UV flux in the hard and soft spectral states, different delays may be expected and hence affecting the observed X-ray/UV relation (see also Section~\ref{subsec:flow}). We note that the \swift/BAT and \maxi\ monitoring light curves show that the outburst of \source\ continued after our pointed XRT and UVOT observations stopped (due to Sun constraints). The X-ray/UV relation may have eventually returned to a positive correlation, tracing out a similar path as the X-ray/nIR relation seen in other black holes \citep[e.g.,][]{coriat2009}. However, it is not clear if the underlying physical processes should be the same.

\section{Discussion}\label{sec:discussion}

\subsection{Inclination and Disk Temperature}
A \chan/HETG spectrum obtained when \source\ was in the soft state appeared featureless, despite the fact that ionized disk winds are expected to be ubiquitous in this spectral state \citep[e.g.,][]{miller2008,miller2012_winds}. If such winds are mostly equatorial, a possible explanation for the lack of ionized absorption features is that \source\ is viewed at relatively low inclination \citep[][]{ponti2012_winds}. This would be consistent with the analysis of disk reflection features, which suggested an inclination of $i \lesssim 20^{\circ}$ \citep[][]{reis2013}. 

Geometrical effects could also account for the fact that the accretion disk of \source\ peaked at $kT_{\mathrm{in}}\simeq$0.5~keV, which is much cooler than typically seen for black hole LMXBs in the soft state \citep[$kT_{\mathrm{in}}\gtrsim$1~keV; e.g.,][]{dunn2011,reynolds2013}. As shown by \citet{munozdarias2013}, disks look cooler when viewed at lower inclination, which can be understood in terms of relativistic effects. \source\ could thus fit into this picture. Alternatively, the low temperature could be the result of a small disk size, e.g., due to a short orbital period (see Section~\ref{subsec:visc}), a truncated disk or a retrograde black hole spin \citep[][]{reis2013}.

\subsection{Viscous Time Scale of the Accretion Disk?}\label{subsec:visc}
Our multi-wavelength light curves revealed a prominent flux dip that first appeared in the optical band, only to be followed by X-rays $\simeq$6~days later. A comparable X-ray lag ($\simeq$8~days) is suggested by cross-correlating the light curves over the full time range covered by our monitoring campaign. A natural explanation for the observed delay may be a (small-scale) mass-transfer instability that originated at the outer edge of the accretion disk and then propagated inward. The time delay could then represent the viscous time scale of the disk, $t_{\mathrm{visc}}(r) = \frac{2}{3\alpha}(\frac{h}{r})^{-2}\Omega_{K}^{-1}$. Here $\alpha$ is the viscosity parameter, $h$ the scale height of the disk, and $\Omega_{K}$ the Keplerian frequency at radius $r$. 

Assuming $\alpha=0.1$, $h/r=0.01$ (i.e., a geometrically thin disk), and $M=8~\Msun$, a viscous time scale of $t_{\mathrm{visc}}=6$~days would yield a disk radius of $r\simeq3400~GM/c^2$ ($\simeq4\times10^9$~cm). This is relatively small, perhaps supporting the fairly short $\simeq$2--4 hr orbital period hinted by its rapid optical flaring (\cite{lloyd2012}, but see \cite{casares2012}). 
On the other hand, a comparable time delay of several days between the nIR and X-rays has also been inferred for black hole LMXBs that have (much) longer orbital periods \citep[e.g., LMC--X3, GX 339--4, and 4U 1957+11;][]{brocksopp2001,homan2005,russell2010}.

The interpretation of the dip as a mass-transfer instability could naturally explain why in all bands the flux recovered to the pre-dip level. However, this picture may not account for the fact that the flux appeared to hit a minimum in the $H$ and $K$ bands later than in the optical/UV (albeit still before the X-rays; Figure~\ref{fig:dip}). This may suggest that multiple (competing) emission mechanisms are operating, or that the flux dip is the result of a different physical process altogether (Section~\ref{subsec:flow}). This is perhaps hinted by the fact that there appears to be a causal connection between the dip and the transition to the soft state $\simeq1-2$~weeks later. This connection would not be obvious to explain in terms of a small-scale mass-transfer instability. 

\subsection{Changes in the Accretion Flow Morphology}\label{subsec:flow}
Interestingly, a very similar flux dip as we observed for \source\ was reported for \gx\ during its 2010 outburst \citep[][]{yan2012}. In that source the UV flux decreased by $\simeq$60\% in $\simeq$2 weeks time, which was associated with a drop in the optical/nIR (by $\simeq$85\%) as well as the radio flux. Within $\simeq$10 days after this drop the source transitioned from a hard to soft state, and the UV flux dip was therefore associated with jet quenching \citep[][]{yan2012}. 

In \source, support for change in accretion morphology (perhaps connected to a jet) is provided the observation that the nIR color suddenly became bluer right around the time of the flux dip (Figure~\ref{fig:lc}). It remained as such until the end of the soft state, which started $\simeq10$~days after the color change. This behavior was not observed in the optical bands, suggesting a relative suppression of the nIR emission during the soft state, i.e., when jets are known to be quenched. 

When the source transitioned back to the hard state, the $K$ and $H$ band flux started to rise and the nIR color become redder again (around day 138 in Figure~\ref{fig:lc}). Although our \smarts\ monitoring stopped at this time, the hinted nIR brightening may be similar to that seen in other black hole LMXBs entering the hard state \citep[e.g.,][]{jain2001,buxton2004,buxton2012,kalemci2005,kalemci2013,russell2010,russell2013,dincer2012,corbel2013}. The optical/nIR secondary maxima seen in these sources have been ascribed to the revival of a jet \citep[e.g.,][]{dincer2012,kalemci2013}. 

Continued \swift\ monitoring revealed that in \source\ the UV emission started to increase $\simeq$20 days after the transition back into the hard state (around day 158 in Figure~\ref{fig:lc}), whereas the X-rays continuously decayed. This caused the relation between the UV and 0.6--10 keV flux to move from a positive correlation (with a slope of $\beta\simeq0.15$) into a negative correlation ($\beta\simeq -0.45$). This turnover in the UV/X-ray relation occurred at $F_X\simeq10^{-9}~\flux$ (0.6--10 keV). The distance toward \source\ is unknown, but if the peak of the outburst (as observed by \maxi) did not exceed the Eddington limit ($L_{\mathrm{EDD}}$), the transition flux corresponds to $\lesssim0.03\times L_{\mathrm{EDD}}$. The UV brightening was accompanied by a dramatic increase in the X-ray hardness ratio, which again suggests a causal connection with changes in the accretion morphology.

This rise in UV emission is reminiscent of the secondary nIR/optical maxima seen in other black hole LMXBs. In a systematic study of a number of different sources and outbursts, \citet{kalemci2013} showed that the nIR peaks $\simeq5-15$~days after the transition from the soft to the hard state. In \source, the UV appears to peak $\simeq10$ days after it starts rising, or $\simeq30$~days after the state transition (around day 168 in Figure~\ref{fig:lc}). This is thus considerably later than the reported nIR peaks \citep[][]{kalemci2013}. In a jet interpretation it might be expected that the UV increases after the nIR \citep[][]{corbel2013,russell2013}, but it is not immediately clear if this delay could be as large as $\simeq20$~days. 

An alternative interpretation for both the multi-wavelength flux dip and the UV peak may be provided in terms of the collapse and recovery of a hot inner flow \citep[][]{veledina2011,veledina2013}. Such a radiatively inefficient flow might replace the inner accretion disk at low mass-accretion rates \citep[e.g.,][]{esin1997}. Analysis of disk reflection features suggested that the disk in \source\ could indeed be subject to evaporation \citep{reis2013}. This model predicts that the outer regions of the hot flow (traced by the optical/nIR emission) collapses several days before an X-ray state transition is observed \citep[causing a decrease in flux;][]{veledina2013}. This is consistent with the time between the dip and the soft state transition observed for \source\ ($\simeq$1--2~weeks). Moreover, the hot flow model predicts that the optical/nIR spectrum hardens (i.e., becomes bluer) in this transition, consistent with the drop in $J-K$ color that we observed before the state transition (Figure~\ref{fig:lc}). 

\citet{veledina2011} interpret the optical/nIR secondary maxima seen after hard state transitions as the recovery of a hot inner flow \citep[see also][]{kalemci2013,veledina2013}. The same mechanism may account for the UV peak (and corresponding X-ray/UV anti-correlation) discussed in this work. However, in the picture of a recovering hot flow, the UV emission should rise before the nIR \citep[][]{veledina2013}, whereas in \source\ there are indications that the nIR started to rise first. Moreover, a delay as long as $\simeq$30 days between the state transition and the UV peak may be difficult to account for \citep[see also][]{kalemci2013}. 

It is worth noting that although UV monitoring is often hampered by substantial interstellar extinction, anti-correlated UV and X-ray emission has also been observed for the neutron star Cyg X-2 \citep[][]{rykoff2010}. On the other hand, such an anti-correlation was not seen for the black hole LMXBs \xte\ and \bhmontse, despite dense UV/X-ray coverage over a large range of fluxes \citep[][]{rykoff2007,armas2012}. The latter source may have never entered a soft state (\cite{armas2012}, see also \cite{corralsantana2013} and \cite{shahbaz2013}), which could possibly explain the absence of an anti-correlation given that the UV rise appears to be connected to a state transition. However, \xte\  was monitored for well over a month after it transitioned from the soft to the hard state, but its X-ray and UV emission remained tightly correlated \citep[][]{rykoff2007}. Future X-ray/UV monitoring efforts of suitable targets are warranted to further understand the role of the UV emission in state transitions.

\acknowledgments
The authors are grateful to the anonymous referee for constructive and insightful comments that helped improve this manuscript. N.D. is supported by NASA through Hubble Postdoctoral Fellowship grant number HST-HF-51287.01-A from the Space Telescope Science Institute. R.C.R. is supported by NASA through the Einstein fellowship program (grant number PF1-120087) and is a member of the Michigan Society of Fellows. J.M.M. gratefully acknowledges support from the Swift guest observer program. N.D. thanks A. Veledina, J. Poutanen, and M. van Dijk for stimulating discussions.

{\it Facilities:} \facility{CXO (HETG), CTIO:1.3m, Swift (XRT,UVOT)}

\bibliographystyle{apj}

\end{document}